\documentclass[aps,pra, twocolumn,groupedaddress]{revtex4-2}

\usepackage{graphicx}
\usepackage{bbold}
\usepackage{dcolumn}% Align table columns on decimal point
\usepackage{bm}% bold math
\usepackage[caption=false]{subfig}
\usepackage{color}

\begin{document}

%\preprint{}

\title{Indefinite causal order in cavity quantum electrodynamics}
%Quantum-controlled order of two Jaynes-Cummings cavities


%\author{}

\author{Luis Octavio Casta\~nos-Cervantes$^1$}
\author{Lorenzo M. Procopio$^{2,3}$}
\email{Contact author: lorenzo.procopio@uni-paderborn.de}
\author{Tim J. Bartley $^{2,3}$}

%\thanks{}
%\affiliation{}
\affiliation{$^{1}$Tecnologico de Monterrey, School of Engineering and Sciences, 14380 Ciudad de Mexico, CDMX, Mexico}
\affiliation{$^{2}$Institute for Photonic Quantum Systems (PhoQS), Paderborn University, Warburger Str. 100, Paderborn, 33098, Germany}
\affiliation{$^{3}$ Department of Physics, Paderborn University, Warburger Str. 100,Paderborn, 33098, Germany}


\date{\today}

\begin{abstract}

Indefinite causal order (ICO) has the potential to be a new resource for quantum information processing. In most of its experiments, ICO has been investigated in a photonic platform. Here we investigate ICO in a cavity quantum electrodynamics (cQED) system composed of two cavities. Our results show that ICO can create highly entangled states of two distant cavity-fields that never interact directly independent of the initial (excited or ground) state of the atom. These entangled states can have the form of one- or two-photon NOON states. We show that ICO can interchange one photon between both cavities without changing the state of the atom, something that is impossible to achieve for two cavities in well-defined order. Furthermore, the vacuum Rabi oscillations either disappear or lose their sinusoidal form with ICO. Our results show the potential that ICO can offer in the paradigm of light-matter interaction for coherently controlling atom-field observables.

\end{abstract}

%\keywords{}

\maketitle

%---------------------------------------> Section
\section{Introduction}

Quantum mechanics allows one to apply quantum operations in a way in which the order of application is indefinite \cite{Chiribella2013, oreshkov2012quantum}. This technique is known as indefinite causal order (ICO) and offers new advantages over existing quantum protocols, comprising operations applied with well-defined order. Those advantages can be found in quantum computation \cite{araujo2014computational,renner2022computational}, communication complexity \cite{guerin2016exponential,feix2015quantum}, quantum communication \cite{ebler2018enhanced, procopio2019communication}, quantum thermodynamics \cite{felce2020quantum,liu2022thermodynamics}, quantum metrology \cite{zhao2020quantum,goldberg2023evading}, and quantum machine learning \cite{ma2024quantum}. Proof-of-principle experimental demonstrations \cite{procopio2015experimental,taddei2021computational,rubino2021experimental,wei2019experimental,nie2020experimental} suggest that ICO has significant potential for quantum information processing tasks. To create ICO, one has to select appropriate degrees of freedom to implement a target system where quantum operations are applied, and a control system which coherently controls the order of those operations. So far, most experimental implementations involving ICO have been limited to photonic platforms using single photons with discrete and continuous variable quantum operations \cite{rozema2024experimental}. Other platforms have been suggested to create ICO using superconducting qubits \cite{felce2021refrigeration} and nuclear magnetic resonance qubits \cite{nie2022experimental}. However, in these implementations, the coherence to control the order is not guaranteed due to the lack of sufficient degrees of freedom to create ICO. This raises the question of whether other platforms might be suitable for implementing ICO, and what other quantum advantages or effects ICO  might offer for those other platforms. 

In this article, we investigate ICO in cavity quantum electrodynamics (cQED), such as in \cite{gleyzes2007quantum}. Unlike the photonic case, the control system is based on the degrees of freedom of a single atom and the quantum operations are physically described by a light-matter interaction model. We use one of the simplest and most fundamental descriptions of coherent radiation–matter models, namely the Jaynes-Cummings model \cite{shore1993jaynes, gerry2023introductory}, but our approach could easily be extended to other interaction models. To the best of our knowledge, the consequences that ICO might have in the paradigm of light-matter interaction have not yet been explored. Ref. \cite{fellous2023comparing} shows that using a light-matter model for the operations, ICO can solve a task with less energetic costs compared to fixed-order scenarios, but it does not explore the effects of ICO \textit{per se} on the atom-field interaction. Here we investigate how ICO impacts atom-field observables and we also show that ICO can create entanglement between two distant cavity-fields that never interact directly independent of the initial (excited or ground) state of the atom and of the time it takes the atom to transit each cavity.  Specifically, one can create Bell-type states for the fields of two spatially separated cavities, and for the case of two cavity-fields initially in the vacuum state, ICO presents an advantage over the fixed-order scenario by being able to generate one-photon NOON states. Recall that a NOON or N00N state for a quantum electromagnetic field with two modes is usually expressed as a superposition of the form $(\vert N \rangle_{1} \otimes \vert 0 \rangle_{2} + e^{iN\theta}\vert 0 \rangle_{1} \otimes \vert N \rangle_{2})/\sqrt{2}$ where $\theta$ is a real number and $\vert n \rangle_{1}\otimes\vert m \rangle_{2}$ represents a state where there are $n$ photons in mode $1$ and $m$ photons in mode $2$ \cite{Noon1,Noon2}. In our case the two modes correspond to the single-modes of the two cavities. Furthermore, we show that ICO can interchange one photon between both cavities without changing the state of the atom, something that is impossible to achieve for two cavities in well-defined order. We also investigate the effect of ICO on the vacuum Rabi oscillations and find that they either disappear or lose their sinusoidal form. Our results show some of the potential that ICO can offer in the paradigm of light-matter interaction. 

Part of the motivation of this work comes from the fascinating advances in atomic interferometry where atomic beam splitters and atomic mirrors can be implemented \cite{AIReview1,AIReview2} producing effects in matter waves similar to those in light. Atom interferometry can in principle be combined with the cQED techniques to implement ICO on new platforms. Especially relevant is the Sagnac atomic interferometer \cite{Sagna2020,Sagna2021,Sagna2024}, where atomic wavepackets are coherently split and then recombined after they have moved along circular trajectories. In principle, this architecture could be used to implement our proposal, since the two cavities could lie along this circumference. For the treatment presented in this article, it is necessary that the internal and the motion degrees of freedom of the atom can be addressed independently.

The article is organized as follows. In Sec.~II the system under consideration is presented and Sec.~III establishes its Hamiltonian. Sec.~IV describes how the system evolves and Secs. V and VI consider the cases of cavities with ICO and in series, respectively. The conclusions are presented in Sec.~VII. 

\section{The system under consideration}

\begin{figure}[htbp]
   \centering
   \subfloat[]{\label{Figure1a}}\includegraphics[scale=0.1]{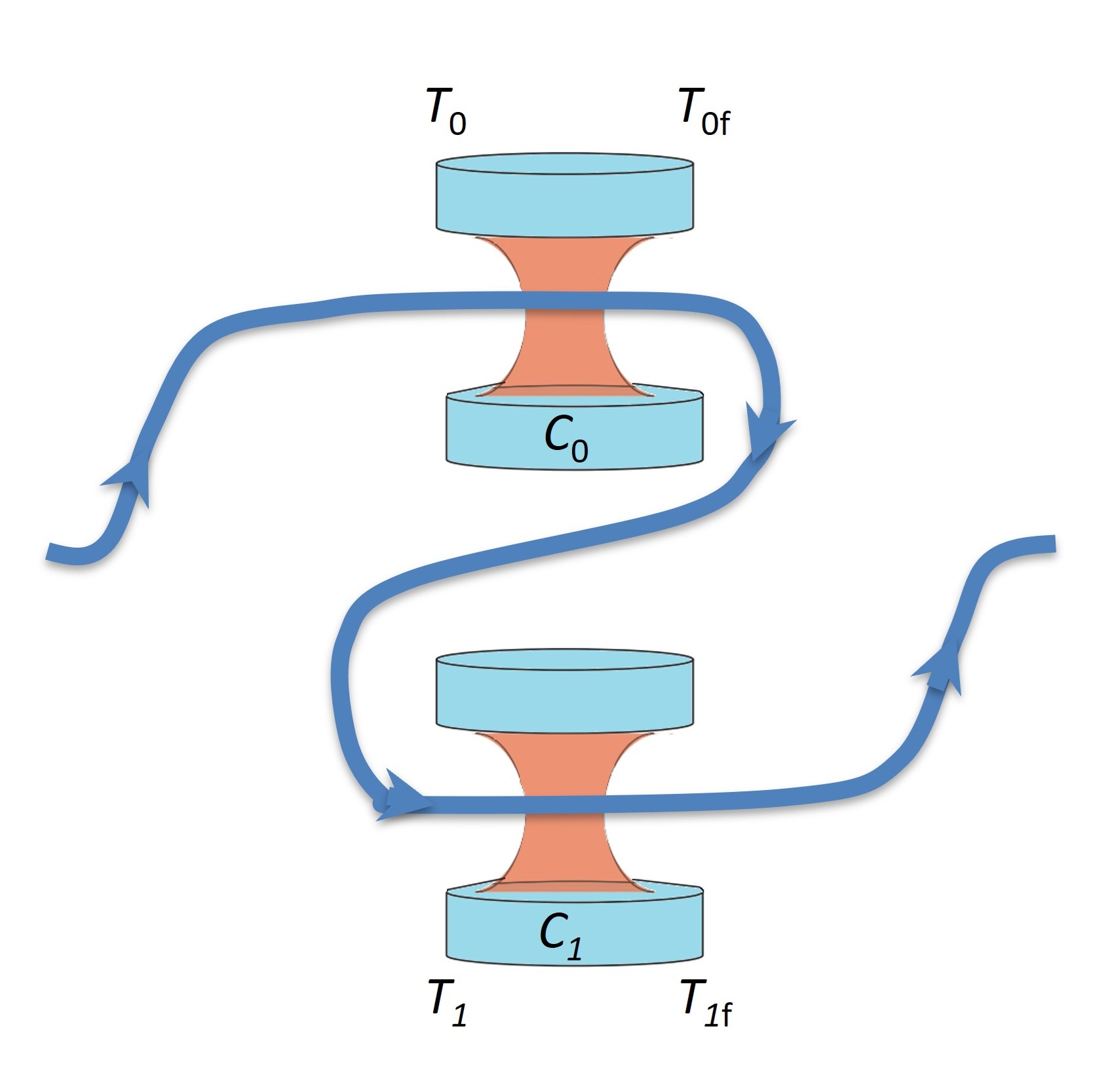} 
   \subfloat[]{\label{Figure1b}}\includegraphics[scale=0.1]{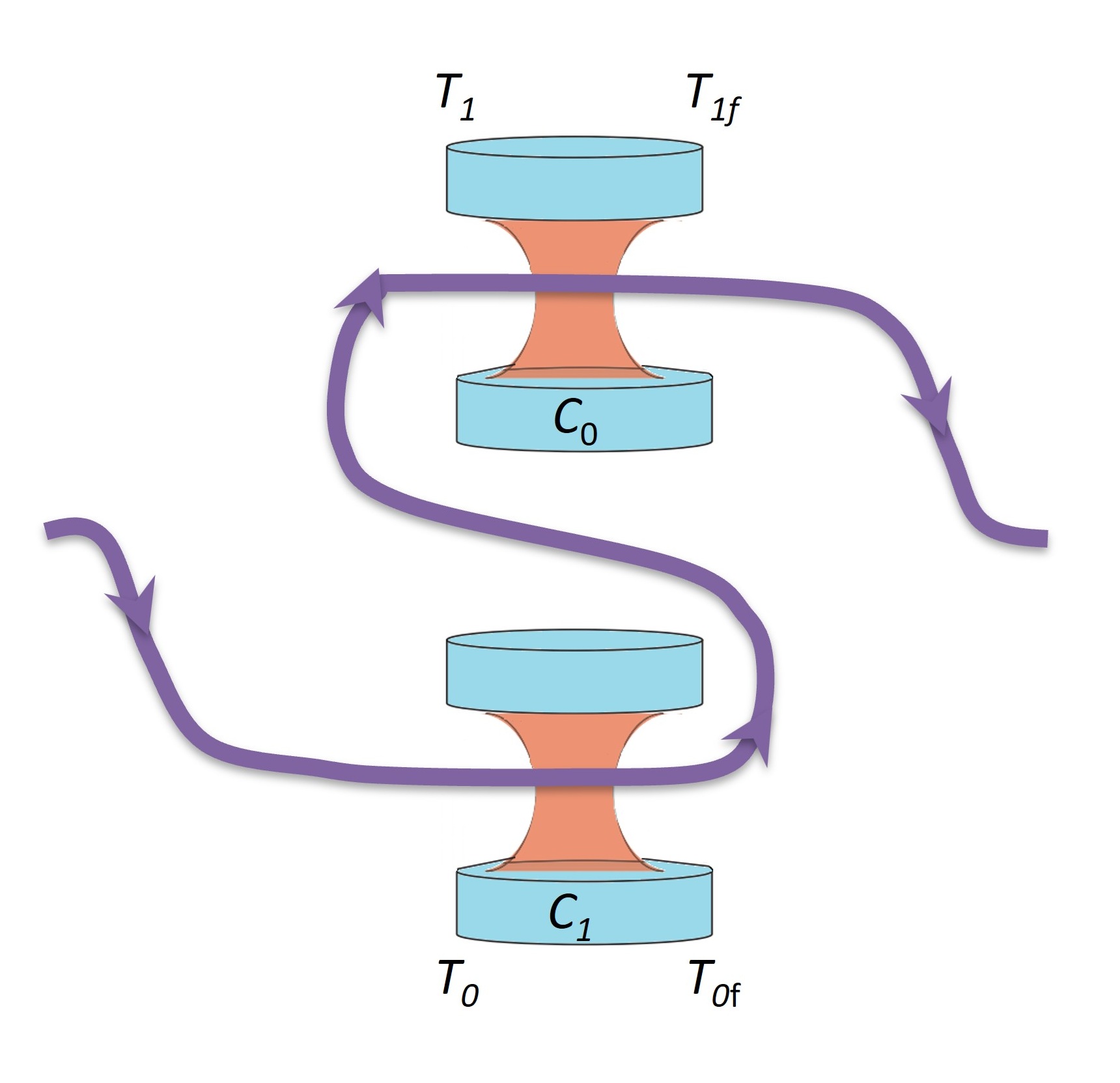} 
   \subfloat[]{\label{Figure1c}}\includegraphics[scale=0.1]{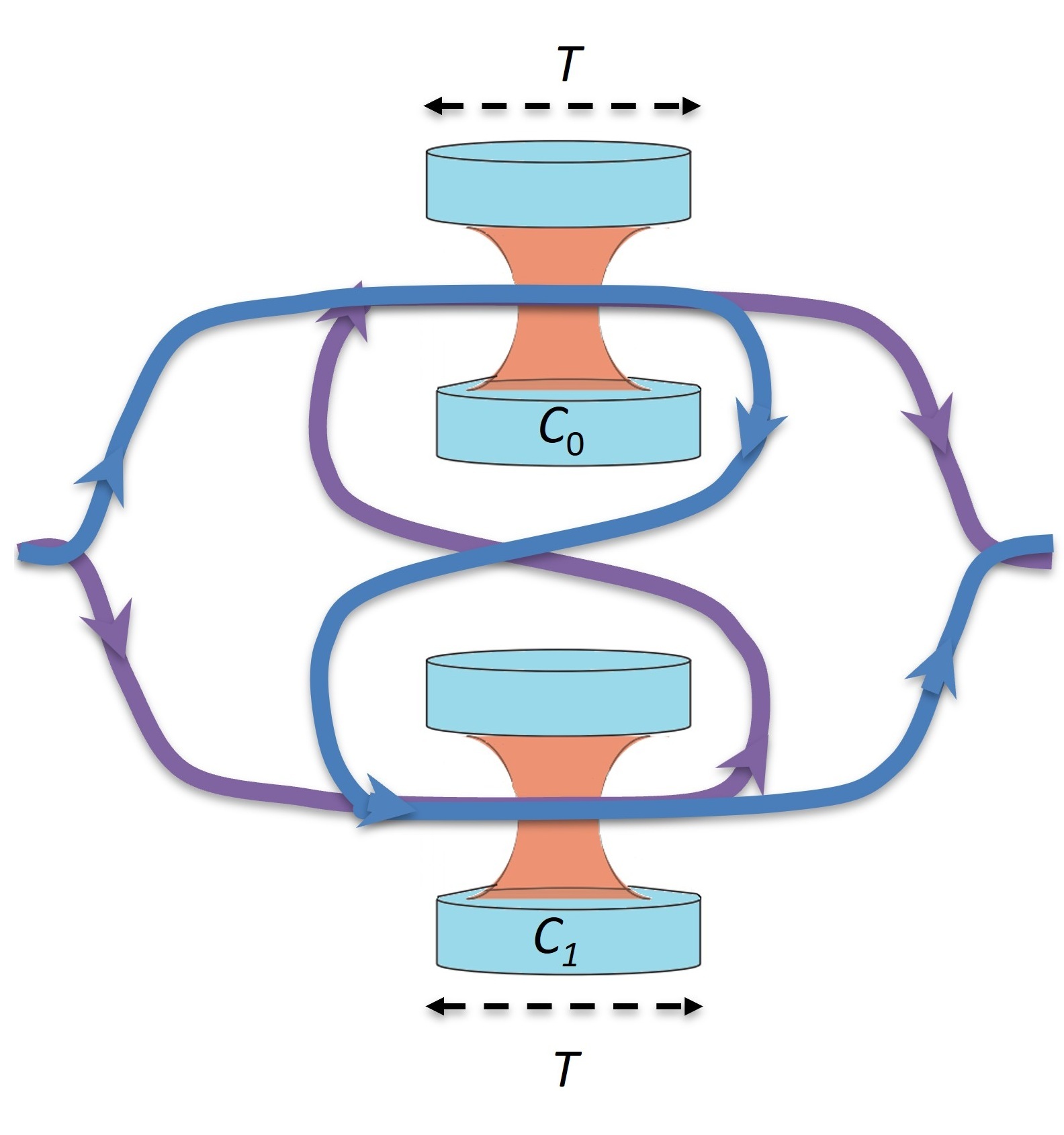} 
   \caption{ICO in a two-cavity system.  In Figs.~\ref{Figure1a} and \ref{Figure1b} the atom traverses the two cavities in the order indicated by the arrows. A control qubit encodes the path: if the control qubit is in the state $\vert 0 \rangle_{c}$ ($\vert 1 \rangle_{c}$), then the atom first traverses cavity $C_0$ ($C_1$) and then cavity $C_1$ ($C_0$). ICO is introduced if the control qubit is in a superposition of its two states, since the atom travels in a superposition of the two paths, see Fig.~\ref{Figure1c}.
} 
  \label{system}
\end{figure}

Fig.~\ref{system} illustrates how ICO can be introduced in a two-cavity system. A single two-level atom passes in succession  through two identical electromagnetic cavities $C_{0}$ and $C_{1}$. When the atom is in the cavity $C_{j}$, it interacts with a single-mode cavity-field with angular frequency $\omega >0$ and with atom-field coupling strength $g$. Once the atom exits both cavities, physical quantities can be measured. The atom can traverse the cavities in the order $C_{0}$ and then $C_{1}$ or viceversa, see Figs.~\ref{Figure1a} and \ref{Figure1b}.
To define the order in which the atom traverses the cavities, we use a control qubit which encodes the path followed by the atom. An orthonormal basis for the state space of the control qubit is $\left\{  \vert 0 \rangle_{c},  \vert 1 \rangle_{c}  \right\}$. If the control qubit is in the state $\vert 0\rangle_{c}$, then the atom will traverse the cavities in the order $C_0$ and then $C_1$. Likewise, if the control qubit is in the state $\vert 1\rangle_{c}$, then the atom will traverse the cavities in the order $C_1$ and then $C_0$. The two orders are shown in Figs.~\ref{Figure1a} and \ref{Figure1b}.  If the control qubit is in a superposition of $\vert 0 \rangle_{c}$ and $\vert 1\rangle_{c}$, then ICO is introduced because the atom will traverse the cavities in a superposition of the two possible paths. This superposition can be created if the atom initially passes through a spatial beamsplitter before entering the cavities \cite{cassettari2000beam}. The most general state of the control qubit is
\begin{equation}
\label{EQcontrol}
\vert \theta,\varphi \rangle_{c} =\cos \theta \vert0\rangle_{c} +e^{i\varphi}\sin \theta \vert1\rangle_{c}	
\end{equation}
\noindent where $\theta \in [ 0, \pi/2 ] $, $\varphi \in [ 0, 2\pi)$. Maximum indefiniteness is achieved when $\theta=\pi/4$ because the control qubit is in an equal probability superposition of the states $\vert 0 \rangle_{c}$ and $\vert 1 \rangle_{c}$. For simplicity, we only consider the case of a resonant interaction, that is, the transition frequency of the atom is   $\omega$. Since the cavities are identical, we also assume that it takes the atom the same time $T>0$ to traverse each cavity.

\section{The Hamiltonian of the system}

The complete system is composed of a two-level atom, two single-mode cavity-fields, and a control qubit. An orthonormal basis for the state space of the atom is $\{ \vert e \rangle, \ \vert g \rangle \}$ where $\vert e \rangle$ is the excited state and $\vert g \rangle$ is the ground state. Meanwhile, an orthonormal basis for the state space of the cavity-field in $C_j$ is $\{ \vert n \rangle_{j} : \ n=0,1,2,... \}$ with $\vert n \rangle_{j}$ a Fock state $(j=0,1)$. The  Hamiltonian of the system is
\begin{equation}
\label{HICO}
	H(t) =  \vert 0 \rangle_{cc} \langle 0 \vert \otimes H_{0}(t) + \vert 1 \rangle_{cc}\langle 1 \vert \otimes H_{1}(t) ,  
\end{equation}
where $H_0 (t)$ ($H_1 (t)$) is the Hamiltonian of the system if the atom first traverses cavity $C_0$ ($C_1$) and then cavity $C_1$ ($C_0$). Explicitly, for $j=0,1$ one has
\begin{equation}
 	H_j(t) = H_{\mathrm{free}} + H^{j}_{\mathrm{int}}(t) ,
\end{equation}
with $H_{\mathrm{free}}$ the free Hamiltonian
\begin{equation}
	H_{\mbox{\tiny free}}= \frac{\hbar \omega}{2} \sigma_{z} + \hbar \omega a_0^{\dagger}a_0 + \hbar \omega a_1^{\dagger}a_1 ,
\end{equation}
and $H_{\mathrm{int}}^{j}(t)$ the Hamiltonian describing the interaction between the atom and each cavity-field
\begin{eqnarray}
	\label{Hin}
	H^{j}_{\mathrm{int}}(t) &=&
	\left\{ 
	\begin{array}{cc}
		\hbar g (\sigma_{-}a^{\dagger}_j + \sigma_{+}a_j) & \mbox{if $T_0\leq t \leq T_{0} + T$}, \cr
		\hbar g (\sigma_{-}a^{\dagger}_{j\oplus 1} + \sigma_{+}a_{j\oplus 1}) & \mbox{if $T_{1} \leq t \leq T_{1} + T$}, \cr
		0 & \mbox{otherwise}.
	\end{array}
	\right. \cr
    &&
\end{eqnarray} 
Here $\oplus$ is the sum modulo 2, $\sigma_z = \vert e \rangle\langle e \vert - \vert g \rangle\langle g \vert$, the transition operators are $\sigma_{-} = \vert g \rangle\langle e \vert$ and $\sigma_{+} = \sigma_{-}^{\dagger}$, $a_{j}$ and $a_{j}^{\dagger}$ are the annihilation and creation operators of the cavity-field in $C_{j}$, $T_{0}$ ($T_{1}$) is the time when the atom enters the first (second) cavity, and $T_{0}+T$ ($T_{1}+T$) is the time when the atom exits the first (second) cavity. Notice that the atom is traveling towards the first cavity between $t=0$ and $t=T_{0}$ and that the atom is traveling from the first cavity towards the second one between $t=T_{0}+T$ and $t=T_{1}$. Also, observe that the interaction of the atom with each cavity field is modeled by the Jaynes-Cummings Hamiltonian \cite{shore1993jaynes,gerry2023introductory} and that $H_{\mathrm{int}}^{j}(t)$ in (\ref{Hin}) has the operators $a_{j}^{\dagger}$ and $a_{j}$ in the first line and $a_{j\oplus 1}^{\dagger}$ and $a_{j\oplus 1}$ in the second line because it describes the fact that the atom first interacts with the cavity-field in $C_{j}$ and then interacts with the cavity-field in $C_{j\oplus 1}$.

In order to simplify the description of the evolution of the system, it is convenient to pass to an interaction picture (IP) using the unitary transformation 
\begin{equation}\label{Ut}
U_I(t)=e^{-  \frac{i}{\hbar} H_{\mathrm{free}} t}.
\end{equation}
If $\vert \Psi(t)\rangle$ is the state of the system at time $t$  in the Schr\"{o}dinger picture (SP), then   
\begin{equation} \label{Tstate}
\vert \Psi_I(t)\rangle=U_I^{\dagger}(t) \vert \Psi(t)\rangle  	
\end{equation}
is the state of the system in the IP. The IP Schr\"{o}dinger equation takes the form
\begin{eqnarray}
\label{SEIP}
\frac{d}{dt}\vert \Psi_I(t)\rangle &=& -\frac{i}{\hbar} \sum_{j=0,1}\vert j \rangle_{cc} \langle j \vert \otimes H^{j}_{\mathrm{int}}(t) \vert \Psi_I(t)\rangle ,
\end{eqnarray}
and the evolution operator $U_{SP} (t,0)$ in the SP is related to the evolution operator $ U_{IP}(t,0)$ in the IP by
\begin{equation}
U_{SP} (t,0)= U_I(t)  U_{IP}(t,0).
\end{equation}
Explicitly, one has
\begin{eqnarray}
	\label{39}
	U_{IP}(t,0) &=& \sum_{j=0,1} \vert j\rangle_{cc}\langle j \vert \otimes \cr
    && 
		\left\{ 
	\begin{array}{cc}
		 \mathbb{I} & \mbox{if $0\leq t < T_{0}$}, \cr
		U_j(t-T_0)   & \mbox{if $T_{0} \leq t \leq T_{0}+T$}, \cr
		U_j(T)    & \mbox{if $T_{0}+T < t < T_{1}$}, \cr
		U_{j\oplus 1}(t-T_1)U_{j}(T) & \mbox{if $T_{1} \leq t \leq T_{1}+T$}, \cr
		U_{j\oplus 1}(T)U_{j}(T)  & \mbox{if $T_{1}+T < t$}. \cr
	\end{array}
	\right. \cr
    &&
\end{eqnarray}
with $\mathbb{I}$ the identity operator and
\[
U_{j}(t)= e^{-i gt (a_{j}^{\dagger}\sigma_{-} + \sigma_{+}a_{j})} .
\]

Using the dressed states of the Jaynes-Cummings Hamiltonian \cite{shore1993jaynes} one can calculate the effect of $U_{j}(t)$ on an arbitrary state of the system. Since we consider the resonant case, the dressed states for the atom and the cavity-field in $C_{j}$ $(j=0,1)$ are given by
\begin{eqnarray}
\label{DS}
\vert \pm ,n \rangle_j &=& \pm \frac{1}{\sqrt{2}} \Bigg( \vert e \rangle \otimes \vert n \rangle_j \pm \vert g \rangle \otimes \vert n+1 \rangle_j \Bigg) ,
\end{eqnarray}
with $n=0,1,2,...$ and satisfy
\begin{eqnarray}
\label{DS2}
\hbar g (a^{\dagger}_j \sigma_{-} + \sigma_{+}a_j) \vert \pm ,n \rangle_j &=&  \pm \hbar g \sqrt{n+1}\vert \pm ,n \rangle_j .
\end{eqnarray}
Together with the ket $\vert g \rangle \otimes \vert 0 \rangle_{j}$, the dressed states in (\ref{DS}) constitute an orthonormal basis for the state space of the subsystem formed by the atom and the cavity-field in $C_{j}$.  Notice that $\hbar g (a^{\dagger}_j \sigma_{-} + \sigma_{+}a_j) \vert g \rangle \otimes \ \vert 0 \rangle_j =  0 .
$

Now, define the excitation number operator
\begin{eqnarray}\label{Ne}
	\mathbb{N}_e &=& \frac{1}{2} (\sigma_z+1)+ a_0^{\dagger}a_0 +a_1^{\dagger}a_1 
		= \frac{1}{\hbar \omega}  H_{\mathrm{free}} + \frac{1}{2} . \quad\quad
\end{eqnarray}
If the atom-cavity-fields subsystem is in the state $\vert l \rangle \otimes \vert n \rangle_{0} \otimes \vert m \rangle_{1}$ with $l=e$ or $g$, then $\mathbb{N}_{e}$ counts the number of excitations because 
\begin{eqnarray}
\label{Ne2}    
\mathbb{N}_{e}\vert l\rangle \otimes \vert n \rangle_{0} \otimes \vert m \rangle_{1} = (\delta_{le}+ n+m) \vert l \rangle \otimes \vert n \rangle_{0} \otimes \vert m \rangle_{1}. \quad\quad
\end{eqnarray}
Since
$H_{\mathrm{free}}$ commutes with both $H_{0}(t)$ and $H_{1}(t)$, it follows that $H_{\mathrm{free}}$ also commutes with $H(t)$. Therefore, $\mathbb{N}_{e}$ is a constant of the motion for $H_{0}(t)$, $H_{1}(t)$, and $H(t)$. Notice that $\mathbb{N}_{e}$ is essentially the same constant of the motion of the Jaynes-Cummings Hamiltonian (in the Jaynes-Cummings model it is a constant of the motion in the most general context where the atom-field interaction is not resonant). Finally, using (\ref{Ne}) one can express the unitary transformation $U_{I}(t)$ in terms of $\mathbb{N}_{e}$:
\begin{eqnarray}
\label{Ne3}
U_{I}(t) &=& e^{-i\omega t (\mathbb{N}_{e} -\frac{1}{2})} .
\end{eqnarray}

\section{The state of the system}

Assume that the initial state of the system is 
\begin{equation}
\label{Psi0}
	\label{Psi0}
	\vert \Psi (0) \rangle = \vert \theta,\varphi \rangle_{c} \otimes \vert a  \rangle \otimes \vert n,m \rangle ,
\end{equation}
where $\vert \theta,\varphi \rangle_{c}$ is the state of the control qubit defined in (\ref{EQcontrol}), $\vert n,m \rangle = \vert n \rangle_{0} \otimes \vert m \rangle_{1}$ with $m$ and $n$ nonnegative integers is a tensor product of Fock states describing the state of the cavity-fields, and $\vert a  \rangle$ is an arbitrary state of the atom. Omitting a global phase, one has
\begin{equation}
\label{Atom}
	\vert a  \rangle = \mathrm{cos} (\xi) \vert e \rangle + e^{i \chi} \mathrm{sin} (\xi) \vert g \rangle ,
\end{equation}
where  $\xi \in [ 0, \pi/2 ] $ and $\chi \in [ 0, 2\pi)$. Observe that the initial state of the atom is $\vert e \rangle$ if $\xi = 0$ and $\chi = 0$, $\vert g \rangle$ if $\xi = \pi/2$ and $\chi =0$, or $\vert \pm \rangle = (\vert e\rangle \pm \vert g \rangle)/\sqrt{2}$ if $\xi = \pi/4$ and $\chi = 0$ for $\vert + \rangle$ or $\chi = \pi$ for $\vert - \rangle$.

Applying the evolution operator in (\ref{39}) and using the dressed states in (\ref{DS}), it follows that the state of the system at time $t \geq T_{1}$ in the IP is given by 
\begin{eqnarray}
\label{CQDstate}
\vert \Psi_I(t) \rangle &=& \mathrm{cos} (\theta) \vert 0 \rangle_{c} \otimes \vert C_1 C_0 (\tau) \rangle \cr
&& + e^{i \varphi}\mathrm{sin} (\theta)\vert 1 \rangle_{c} \otimes	\vert C_0 C_1 (\tau) \rangle ,
\end{eqnarray}
where 
\begin{eqnarray}
\label{tau}
\tau &=& \left\{
\begin{array}{cc}
t-T_1 & \mbox{if $T_{1} \leq t \leq T_{1} + T$}, \cr
T & \mbox{if $T_{1} + T < t$}.
\end{array}
\right.  
\end{eqnarray}
and $\vert C_i C_j (\tau) \rangle$ is the state of the atom-cavity-fields subsystem when the atom first traverses cavity $C_j$ and then cavity $C_i$. Explicitly, one has
\begin{eqnarray}
\label{orderc1c0}
\vert C_{1}C_{0}(\tau) \rangle
&=&
\vert e \rangle \otimes \vert \phi_{e}(\tau) \rangle
+ \vert g \rangle \otimes \vert \phi_{g}(\tau) \rangle , 
\end{eqnarray}
and
\begin{eqnarray}
\label{orderc0c1}
	\vert C_0C_1 (\tau) \rangle &=& 
\vert e \rangle \otimes \vert \Phi_{e}(\tau) \rangle
+ \vert g \rangle \otimes \vert \Phi_{g}(\tau) \rangle , 
\end{eqnarray}
where
\begin{eqnarray}
\label{CS2}
\vert \phi_{e}(\tau) \rangle &=& \Big[ c_{1}\vert n \rangle_{0} + c_{2}\vert n-1 \rangle_{0} \Big] \otimes \vert m \rangle_{1} \cr
&& + \Big[ c_{5}\vert n \rangle_{0} + c_{6}\vert n+1 \rangle_{0} \Big] \otimes \vert m-1\rangle_{1} , \cr
\vert \phi_{g}(\tau) \rangle &=& \Big[ c_{7}\vert n \rangle_{0} + c_{8}\vert n+1 \rangle_{0} \Big] \otimes \vert m \rangle_{1} \cr
&& + \Big[ c_{3}\vert n \rangle_{0} + c_{4}\vert n-1 \rangle_{0} \Big] \otimes \vert m+1\rangle_{1} , 
\end{eqnarray}
and
\begin{eqnarray}
\label{CS2b}
\vert \Phi_{e}(\tau) \rangle &=& \vert n \rangle_{0} \otimes \Big[ s_{1}\vert m \rangle_{1} + s_{2}\vert m-1 \rangle_{1} \Big] \cr
&& + \vert n-1 \rangle_{0} \otimes \Big[ s_{5}\vert m \rangle_{1} + s_{6}\vert m+1 \rangle_{1} \Big] , \cr
\vert \Phi_{g}(\tau) \rangle &=& \vert n+1 \rangle_{0} \otimes \Big[ s_{3}\vert m \rangle_{1} + s_{4}\vert m-1 \rangle_{1} \Big] \cr
&& + \vert n \rangle_{0} \otimes \Big[ s_{7}\vert m \rangle_{1} + s_{8}\vert m+1 \rangle_{1} \Big] .
\end{eqnarray}
Here $\vert l,n,m \rangle = \vert l \rangle \otimes \vert n \rangle_{0} \otimes \vert m \rangle_{1}$ with $l=e,g$ and the coefficients $c_{j} = c_j (\tau)$ and $s_{j} = s_j (\tau)$ depend on $\tau$ and are given in the Appendix, see  (\ref{Cterms}) and (\ref{Sterms}). The dependence of $c_{j}$ and $s_{j}$ on $\tau$ was omitted in (\ref{CS2}) to have more succinct expressions. Note that $\vert C_i C_j (\tau) \rangle$  depends on the parameters $n$, $m$, $g$, $T$, $\xi$, and $\chi$ through the coefficients. In addition, note that $\vert C_{i}C_{j}(\tau) \rangle$ is a normalized ket that is a linear combination of pairwise orthogonal kets, that is,  $ \langle C_1 C_0 (\tau) \vert C_1 C_0 (\tau) \rangle = \langle C_0 C_1 (\tau)\vert C_0 C_1 (\tau) \rangle=1$. 

An alternative expression of $\vert \Psi_{I}(t) \rangle$ that is useful to identify the interference effects due to ICO is obtained by writing  $\vert \Psi_{I}(t) \rangle$ in terms of the orthonormal control qubit states $\vert \theta , \varphi \rangle_{c}$ defined in (\ref{EQcontrol}) and
\begin{eqnarray}
\label{controlB}
\vert \theta , \varphi \rangle_{c}'
&=& \mathrm{sin}\theta \vert 0 \rangle_{c} - e^{i\varphi} \mathrm{cos}\theta \vert 1 \rangle_{c} .
\end{eqnarray}
One finds that
\begin{eqnarray}
\label{controlB2}
\vert \Psi_{I}(t) \rangle
&=& \vert \theta , \varphi \rangle_{c} \otimes \left[ \mathrm{cos}^{2}\theta \vert C_{1}C_{0}(\tau) \rangle + \mathrm{sin}^{2}\theta \vert C_{0}C_{1}(\tau) \rangle \right]
\cr
&& + \vert \theta , \varphi \rangle_{c}' \otimes \frac{\mathrm{sin}(2\theta)}{2}\Big[ \ \vert C_{1}C_{0}(\tau) \rangle - \vert C_{0}C_{1}(\tau) \rangle \ \Big] . \cr
&&
\end{eqnarray}
Observe that the expression for $\vert \Psi_{I}(t) \rangle$ in (\ref{controlB2}) explicitly shows that the atom-cavity fields subsystem is found in a superposition of the states $\vert C_{1}C_{0}(\tau) \rangle$ and $\vert C_{0}C_{1}(\tau) \rangle$ associated with the two paths available to the atom if one measures the state of the control qubit to see if it is found in the state $\vert \theta , \varphi \rangle_{c}$ or $\vert \theta , \varphi \rangle_{c}'$.

Before continuing it is important to point out that one cannot observe interference phenomena due to ICO in the atom-cavity-fields subsystem from the state $\vert \Psi_{I}(t) \rangle$ in (\ref{CQDstate}) and (\ref{controlB2}).  This can be seen explicitly by tracing over the degrees of freedom of the control qubit. One obtains that the density operator in the IP describing the state of the atom-cavity-fields subsystem is
\begin{eqnarray}
\label{Traza}
\rho_{I}(t)
&=&
\mathrm{Tr}_{c}\left[ \  \vert \Psi_{I}(t) \rangle\langle \Psi_{I}(t) \vert \ \right] , \cr
&=& \mathrm{cos}^{2}( \theta ) \vert C_{1}C_{0}(\tau) \rangle\langle C_{1}C_{0}(\tau) \vert \cr
&& +
\mathrm{sin}^{2}( \theta ) \vert C_{0}C_{1}(\tau) \rangle\langle C_{0}C_{1}(\tau) \vert ,
\end{eqnarray}
where $\mathrm{Tr}_{c}[ \cdot ]$ denotes the trace over the degrees of freedom of the control qubit. Observe that $\rho_{I}(t)$ simply describes a statistical mixture of the two states $\vert C_{1}C_{0}(\tau)\rangle$ and $\vert C_{0}C_{1}(\tau)\rangle$ associated with the two paths available for the atom, namely, first traversing through cavity $C_{0}$ and then through cavity $C_{1}$ or viceversa. To observe interference phenomena due to ICO one can perform a projective measurement on the control qubit in the basis $\{ \vert \theta, \varphi \rangle_{c} , \ \vert \theta, \varphi \rangle_{c}' \}$. When $\theta = \pi/4$ and $\varphi =0$ this corresponds to the basis $\{ \vert \pm \rangle_{c} = (\vert 0 \rangle_{c} \pm \vert 1 \rangle_{c})/\sqrt{2}\}$.
%%%
%%%
%%%

\section{Effects of ICO}

We now determine the effects of ICO on the system, see Fig.~\ref{Figure1c}. Recall that the state of the complete system is given in (\ref{CQDstate}) and (\ref{controlB2}), expressions that are completely general given the initial state in (\ref{Psi0}) and (\ref{Atom}). In all that follows, we restrict our discussion to the case of maximum indefiniteness, that is, $\theta = \pi /4$ and $\varphi = 0$. In addition, we only consider an initial state of the form (\ref{Psi0})-(\ref{Atom}) and the situation where the atom has exited both cavities, that is, $t> T_{1}+T$ and, consequently, $\tau = T$ [see (\ref{tau})]. 

After the atom exits the two cavities, the atomic wavepackets are coherently recombined in a balanced atomic beamsplitter to erase the information about the path followed by the atom. This operation is represented by applying a Hadamard transformation $\mathcal{H}$ on the control qubit at a time $t > T_{1}+T$, where
\[
\mathcal{H} \vert j \rangle_{c} = \frac{1}{\sqrt{2}}\left[ \vert 0 \rangle_{c} + (-1)^{j}\vert 1 \rangle_{c} \right] \quad\quad (j=0,1).
\]
Note that (\ref{CQDstate}) is the state of the system at time $t$ in the IP and that the Hadamard transformation must be applied to the state of the system in the SP. The state of the system in the SP at time $t > T_{1}+T$ immediately after applying the Hadamard transformation at time $t$ is 
\begin{eqnarray}
\label{controlB3}
\vert \psi(t) \rangle &=& \mathcal{H} U_I(t)  \vert \Psi_I(t) \rangle , \cr
&=&
U_{I}(t) \Bigg\{
\vert 0 \rangle_{c} \otimes \frac{1}{2} \Big[ \ \vert C_{1}C_{0}(T)\rangle + \vert C_{0}C_{1}(T)\rangle \ \Big] \cr
&& 
\ + \vert 1 \rangle_{c} \otimes \frac{1}{2} \Big[ \ \vert C_{1}C_{0}(T)\rangle - \vert C_{0}C_{1}(T)\rangle \ \Big]
\Bigg\} .
\end{eqnarray}

If one makes a measurement of the state of the control qubit immediately after applying the Hadamard transformation to see if it is found in the state $\vert 0 \rangle_{c}$ or the state $\vert 1 \rangle_{c}$ (in the jargon of optics one would say that \textit{one detects the atom in mode} $j$), then the probability to find the control qubit in the state $\vert j \rangle_{c}$ is $\mathcal{N}_{j}^{2}$ with
\begin{equation}
\mathcal{N}_{j}= \frac{1}{\sqrt{2}}\sqrt{  1+ (-1)^{j} \mathrm{Re}\Bigl[ \langle C_1C_0 (T) \vert C_0C_1 (T) \rangle \Bigr] } ,
\end{equation}
and the state of the system immediately after the measurement is
\begin{eqnarray}
\label{CQDfinal}
	\vert \psi_{j}(t) \rangle_{} &=& \frac{1}{\mathcal{N}_{j}} \vert j \rangle_{cc} \langle j\vert \mathcal{H}   U_I(t) \vert \Psi_I(t) \rangle , \cr 
	&=& \vert j \rangle_{c} \otimes\frac{ U_I(t)}{2 \mathcal{N}_{j}} 	  \Bigl[ \vert C_1 C_0 (T) \rangle + (-1)^{j}\vert C_0C_1 (T)\rangle \Bigr]  , \cr
    &&
\end{eqnarray}
if one finds the control qubit in the state $\vert j \rangle_{c}$ $(j=0,1)$. Notice that $\mathcal{N}_{j}$ is the normalization constant and that 
\begin{eqnarray}
&& \langle C_1C_0 (T) \vert C_0C_1 (T) \rangle \cr
&=& c_1(T)^* s_1(T) + c_2(T)^* s_5(T)+ c_3(T)^*s_8(T) \cr
&& + c_5(T)^*s_2(T) + c_7(T)^*s_7(T) + c_8(T)^* s_3(T). \quad\quad
\end{eqnarray}
Recall that $c_{j}(T)$ and $s_{k}(T)$ are defined in the Appendix. 

Observe that one has
\begin{eqnarray}
\label{Uno2}
\mathcal{N}_{j}^{2} = 0 \quad &\Leftrightarrow& \quad \mathrm{Re}\Big[ \langle C_{1}C_{0}(T) \vert C_{0}C_{1}(T) \rangle \Big] = (-1)^{j+1}, \cr
&\Leftrightarrow& \quad
\vert C_{1}C_{0}(T) \rangle = (-1)^{j+1} \vert C_{0}C_{1}(T) \rangle .
\end{eqnarray}
We have used the Cauchy-Schwarz inequality in the second line of (\ref{Uno2}). When (\ref{Uno2}) is satisfied $(j=0,1)$, it follows from (\ref{controlB3}) that the state of the system in SP immediately before the measurement of the state of the control qubit is (except, possibly, for a global phase factor) 
\begin{eqnarray}
\label{Uno3}
\vert \psi (t)\rangle
&=&
\vert j \oplus 1 \rangle_{c} \otimes U_{I}(t)\vert C_{1}C_{0}(T) \rangle ,
\end{eqnarray}
so there is no interference due to ICO and the system behaves as if the cavities were in series (recall that $\oplus$ means sum modulo 2). In all that follows we assume that $\mathcal{N}_{j}^{2} >0$ when discussing that the control qubit was found in the state $\vert j \rangle_{c}$. An especially relevant situation is when both cavity-fields are initially in their respective vacuum states ($n=m=0$). In this case the probability $P_{j\vert l}$ to find the control qubit in the state $\vert j \rangle_{c}$ given that the atom is initially in the state $\vert l \rangle$ with $j=0$ or $1$ and $l = e$, $g$, or $\pm$ is given by 
\begin{eqnarray}
\label{ProbControl}
P_{0\vert e} &=& \frac{1}{2}\left[ 1 + \mathrm{cos}^{4}(gT) + \mathrm{sin}(gT)\mathrm{sin}(2gT) \right] \ > \ 0.15, \cr
P_{0\vert g} &=& 1, \cr
P_{0\vert \pm} &=& \frac{1}{4}\left[ 3 + \mathrm{cos}^{4}(gT) + \mathrm{sin}(gT)\mathrm{sin}(2gT) \right] \ > \ 0.57, \cr
P_{1\vert e} &=& 2\mathrm{sin}^{4}\left( \frac{gT}{2} \right) \mathrm{sin}^{2}(gT) , \cr
P_{1\vert g} &=& 0, \cr
P_{1\vert \pm} &=& \mathrm{sin}^{4}\left( \frac{gT}{2} \right) \mathrm{sin}^{2}(gT) .
\end{eqnarray}
The numbers appearing on the righthand side of the first and third equations are lower bounds for the corresponding probabilities. Here $P_{1\vert l} = 1- P_{0\vert l}$. Notice that $P_{0\vert l}$ is always positive and that $P_{1\vert l} = 0$ ($l=e$ or $\pm$) if and only if $gT = N\pi$ with $N$ a positive integer. In these cases one can adjust the transit time $T$ so that the control qubit is always found in the state $\vert 0 \rangle_{c}.$

The \textit{erasing of the which-path-information of the atom} performed by the Hadamard transformation consists in obtaining a state where two different superpositions of $\vert C_{0}C_{1}(T)\rangle$ and $\vert C_{1}C_{0}(T)\rangle$ appear, see (\ref{controlB3}). One then obtains one superposition or the other depending on the result of the measurement of the state of the control qubit to see if it is found in $\vert 0\rangle_{c}$ or $\vert 1\rangle_{c}$. The same result is obtained if one expresses $\vert \Psi_{I} (t) \rangle$ in the form (\ref{controlB2}) with $\theta = \pi/4$ and $\varphi = 0$, one passes to the SP  [compare (\ref{controlB2}) and (\ref{controlB3})], and then one measures the state of the control qubit to see if it is found in one of the orthonormal kets $\vert \pm \rangle_{c}$.

%%%-------------------------------
%%%-------------------------------
%%%-------------------------------

\subsection{Entanglement between the cavity fields}

The state of the system in SP immediately after the measurement of the control qubit is given in (\ref{CQDfinal}). Using (\ref{Ne3}), (\ref{orderc1c0})-(\ref{CS2b}) it can be expressed as
\begin{eqnarray}
\label{EMC}
\vert \psi_{j} (t) \rangle
&=& \vert j \rangle_{c} \otimes  \frac{e^{-i\omega t (n+m - \frac{1}{2})}}{2\mathcal{N}_{j}}\sum_{l = e,g} \vert l \rangle \otimes \mathcal{N}_{lj}\vert F_{lj} (t) \rangle , \cr
&&
\end{eqnarray}
where $j\in \{ 0,1 \}$ and the normalized cavity-fields state $\vert F_{lj} (t) \rangle$ is given by

\begin{widetext}
\begin{eqnarray}
\label{EdoCampos}
\vert F_{ej} (t) \rangle 
&=& \ \frac{1}{\mathcal{N}_{ej}}\Bigg( \Big\{ e^{-i\omega t} \left[ c_{1} + (-1)^{j}s_{1} \right] \vert n \rangle_{0} + \left[ c_{2} + (-1)^{j}s_{5} \right] \vert n-1 \rangle_{0} \Big\} \otimes \vert m \rangle_{1} \cr
&& + \Big\{ \left[ c_{5} + (-1)^{j}s_{2} \right] \vert n \rangle_{0}  + e^{-i\omega t} c_{6}\vert n+1 \rangle_{0} \Big\} \otimes \vert m-1 \rangle_{1}  + (-1)^{j}e^{-i\omega t} s_{6}\vert n-1 \rangle_{0} \otimes \vert m+1 \rangle_{1} \Bigg) ,
\end{eqnarray}
and
\begin{eqnarray}
\label{EdoCamposb}
\vert F_{gj} (t) \rangle 
&=& \ \frac{1}{\mathcal{N}_{gj}}\Bigg( \Big\{ \left[ c_{7} + (-1)^{j}s_{7} \right] \vert n \rangle_{0} + e^{-i\omega t}\left[ c_{8} + (-1)^{j}s_{3} \right] \vert n+1 \rangle_{0} \Big\} \otimes \vert m \rangle_{1}  \cr
&& + \Big\{ e^{-i\omega t} \left[ c_{3} + (-1)^{j}s_{8} \right] \vert n \rangle_{0}  + c_{4}\vert n-1 \rangle_{0} \Big\} \otimes \vert m+1 \rangle_{1} + (-1)^{j}s_{4}\vert n+1 \rangle_{0} \otimes \vert m-1 \rangle_{1} \Bigg) .
\end{eqnarray}
\end{widetext}

In these expressions the $c_{k}$ and $s_{k}$ are evaluated at time $T$. This dependence has been omitted to have more succinct expressions. The normalization constants $\mathcal{N}_{lj}$ ($l=e,g$) are  presented in (\ref{Nlj}) of the Appendix. Notice that it is easy to take the partial trace of the density operator associated with $\vert F_{lj}(t) \rangle$ with respect to the degrees of freedom of the cavity-field in $C_{1}$ using (\ref{EdoCampos}) and (\ref{EdoCamposb}). We emphasize that the expressions in (\ref{EMC})-(\ref{EdoCamposb}) hold for any initial state of the form (\ref{Psi0})-(\ref{Atom}). Choosing $\xi$ and $\chi$ appropriately [see paragraph after (\ref{Atom})] one can consider any initial state for the atom.

Assume that, immediately after the measurement of the state of control qubit, one measures the state of atom to see if it is found in the state $\vert e \rangle$ or $\vert g \rangle$. Then, it follows from (\ref{EMC}) that the state of the system immediately after the measurement of the state of the atom is
\begin{eqnarray}
\label{EdoFinal}
\vert \psi_{lj}(t) \rangle &=& \vert j \rangle_{c} \otimes \vert l \rangle \otimes \vert F_{lj} (t) \rangle ,
\end{eqnarray}
if the atom was found in the state $\vert l \rangle$ with $l = e$ or $g$. Hence, the cavity-fields are in the state $\vert F_{lj} (t) \rangle$. 

We now characterize the entanglement between the cavity-fields. We single-out the situation where the cavity-fields are initially in their respective vacuum states because it is the more experimentally accessible case \cite{gleyzes2007quantum}.

To quantify the entanglement we use the linear entropy. Recall the following properties \cite{Breuer} of the linear entropy $S_{L}(\rho) = 1 -\mathrm{Tr}(\rho^{2})$ of a density operator $\rho$ where $\mathrm{Tr}$ denotes the trace.  Assume that $\rho$ is a pure density operator of a bipartite quantum system $\mathcal{A}+\mathcal{B}$. The reduced density operators of $\mathcal{A}$ and $\mathcal{B}$ are respectively given by $\rho_{A} = \mathrm{Tr}_{B}(\rho)$ and $\rho_{B} = \mathrm{Tr}_{A}(\rho)$ where $\mathrm{Tr}_{A}$ and $\mathrm{Tr}_{B}$ denote the partial traces with respect to the degrees of freedom of $\mathcal{A}$ and $\mathcal{B}$, respectively. Then, $S_{L}(\rho_{A}) = S_{L}(\rho_{B})$. Also, $\rho$ is an entangled state if and only if $S_{L}(\rho_{A})>0$. If the state spaces of $\mathcal{A}$ and $\mathcal{B}$ have finite dimensions $d_{A}$ and $d_{B}$, respectively, then $S_{L}(\rho_{A}) \leq 1 - 1/\mathrm{min}\{ d_{A}, d_{B} \}$. In particular, $\rho$ is a maximally entangled state when $S_{L}(\rho_{A})$ is equal to the upper bound. Whenever we refer to the linear entropy of a cavity-fields state $\vert F \rangle$ it is to be understood that it is the linear entropy of the associated reduced density operator $\mathrm{Tr}_{j}(\vert F \rangle\langle F \vert)$ with $\mathrm{Tr}_{j}$ the trace with respect to the degrees of freedom of the cavity-field in $C_{j}$ ($j = 0$ or $1$), so that the linear entropy quantifies the entanglement of the state $\vert F \rangle$.

Since the cavity-fields are in the state $\vert F_{lj} (t) \rangle$, it follows that the subspace spanned by $\beta_{F} = \{ \vert n_{0},m_{1} \rangle: \ n_{0} = n, n\pm 1 \ \mathrm{and} \ m_{1}=m, m\pm 1 \}$ is invariant under the pure state $\vert F_{lj} (t) \rangle\langle F_{lj} (t) \vert$, so one can consider that $\vert F_{lj} (t) \rangle\langle F_{lj} (t) \vert$ acts in a bipartite state space of dimension $3^{2}$. In this case, the maximum value of the linear entropy is $2/3$. The maximum value of the linear entropy can be reduced if one can consider a set $\beta_{F}$ with less elements. For example, the maximum value of the linear entropy is $1/2$ if $\beta_{F}$ contains only $2^{2}$ elements as in the case of the \textit{Bell-like} states defined below. 

In order to express the results succinctly, we consider three types of \textit{Bell-like} states:  
\begin{enumerate}
\item The cavity-fields \textit{swap} one photon
\begin{eqnarray}
\label{swap}
\vert S\pm \rangle &=& \frac{1}{\sqrt{2}} \Big( \vert n+1,n-1 \rangle \pm \vert n-1,n+1 \rangle \Big) .
\end{eqnarray}
These states require $n\geq 1$. When $n=1$, $\vert S\pm \rangle$ are two-photon NOON states.
\item The cavity-fields \textit{gain} one photon
\begin{eqnarray}
\label{gain}
\vert G\pm \rangle &=& \frac{1}{\sqrt{2}} \Big( \vert n,n+1 \rangle \pm \vert n+1,n \rangle \Big) ,
\end{eqnarray}
These states require $n\geq 0$ and $\vert G\pm\rangle$ are one-photon NOON states when $n=0$.
\item The cavity-fields \textit{lose} one photon
\begin{eqnarray}
\label{lose}
\vert L\pm \rangle &=& \frac{1}{\sqrt{2}} \Big( \vert n-1,n \rangle \pm \vert n,n-1 \rangle \Big) .
\end{eqnarray}
These states require $n\geq 1$ and $\vert L_{\pm}\rangle$ are one-photon NOON states when $n=1$.
\end{enumerate}
These are referred to as \textit{Bell-like} states because one can think of the kets $\vert n \rangle_{j}$ and $\vert m \rangle_{j}$ ($j=0,1$) appearing in the states $(1/\sqrt{2})(\vert n,m\rangle \pm \vert m,n\rangle)$ as two orthonormal states of a qubit. In this sense, they are maximally entangled states and their linear entropy is $1/2$.

Consider the case where the two cavity-fields initially have the same number of photons, that is, they are initially in the state $\vert n,n\rangle$ with $n$ a nonnegative integer. Tables 1-3 present (except, possibly, for a global phase factor) the state $\vert F_{lj}(t) \rangle$ of the cavity-fields subsystem after the measurement of the state of the atom. Table 1 considers $n=0$, while Tables 2 and 3 have $n>0$. The first column of each table from left to right shows the initial state of the atom, while the first row of each table from top to bottom indicates the state in which the control qubit and the atom were found after their respective measurements. Also, NP stands for \textit{not possible} because the probability for such a case is zero. In addition, Table 2 (3) considers the case where the control qubit was found in the state $\vert 0 \rangle_{c}$ ($\vert 1 \rangle_{c}$). The scalars $d_{ij}$ and the probabilities associated with each situation are presented in (\ref{abdc})-(\ref{PC}) of the Appendix. All the coefficients must be evaluated at $T$ with the values of $\chi$ and $\xi$ corresponding to the considered case, see the paragraph below (\ref{Atom}).

Let us analyze some interesting cases. Assume that the cavity-fields are initially in the state $\vert 0, 0 \rangle$. Table 1 shows the different possible results with $\vert \phi_{gg}^{0} \rangle$ defined in Table 2. Especially relevant is the case where the initial state of the atom is $\vert e \rangle$ because the cavity-fields are prepared either in the maximally entangled one-photon NOON states $\vert G\pm \rangle$ or in the separable state $\vert 0,0\rangle$. One can exclude de latter possibility by adjusting the transit time $T$ to satisfy $gT = (2N-1)\pi/2$ with $N$ a positive integer [see (\ref{PC}) in the Appendix]. From (\ref{ProbControl}) observe that, for this transit time $T$, the probability to find the control qubit in the state $\vert j \rangle_{c}$ ($j=0,1$) given that the initial state of the atom was $\vert e \rangle$ is given by $P_{j\vert e} = 1/2$, so that it is equally probable for the cavity-fields to be prepared in one of the states $\vert G\pm \rangle$.

Now assume that the cavity-fields are initially in the state $\vert n, n \rangle$ with $n$ a positive integer. Tables 2 and 3 present the possible states for the cavity-fields if the control qubit was found in the state $\vert 0 \rangle_{c}$ or $\vert 1 \rangle_{c}$, respectively. The kets $\vert \phi_{ll}^{0} \rangle$ $(l=e,g)$ are defined in Table 2.

Observe from Table 2 that the cavity-fields can be prepared in one of the highly entangled states $\vert S+\rangle$, $\vert G+\rangle$, or $\vert L+ \rangle$. By adjusting the transit time $T$ one can exclude the possibility of obtaining the separable state $\vert n,n \rangle$. If the atom is initially in the excited state $\vert e \rangle$ and $gT\sqrt{n +1} = (2N-1)\pi /2$ with $N$ a positive integer, then one can guarantee that the cavity-fields are prepared in either $\vert S+ \rangle$ (if the atom is found in the state $\vert e \rangle$) or $\vert G+ \rangle$ (if the atom is found in the state $\vert g \rangle$), see (\ref{abdc}) and (\ref{d0j}). In the former case, the atom simply acts as an intermediary that allows the cavity-fields to interchange a photon because the cavity-fields are prepared in the $\vert S+\rangle$ state and the atom \textit{starts and is found} in the excited state. Similarly, if the atom is initially in the ground state $\vert g \rangle$ and $gT\sqrt{n} = (2N-1)\pi /2$ with $N$ a positive integer, then one can guarantee that the cavity-fields are prepared in either $\vert L+ \rangle$ or $\vert S+ \rangle$ if the atom is found in the state $\vert e \rangle$ or $\vert g \rangle$, respectively, see (\ref{abdc}) and (\ref{d0j}). In the latter case the atom acts as an intermediary that allows the cavity-fields to interchange a photon because the cavity-fields are prepared in the state $\vert S+\rangle$ and the atom \textit{starts and is found} in the ground state. Finally, if the initial state of the atom is $\vert \pm \rangle$, then the cavity-fields are left in a superposition of the states corresponding to the cases where the initial state of the atom is $\vert e\rangle$ or $\vert g \rangle$. There are transit times $T$ such that the linear entropy of this superposition is $2/3$. Also, choosing $gT\sqrt{n +1} = (2N-1)\pi /2$ or $gT\sqrt{n} = (2N-1)\pi /2$ with $N$ a positive integer, guarantees that $\vert \phi_{ee}^{0} \rangle $ or $\vert \phi_{gg}^{0} \rangle$ reduce to $\vert S+ \rangle$, respectively.

It should also be noted from Table 3 that, if the control qubit was found in the state $\vert 1 \rangle_{c}$, then the cavity-fields are always prepared in a highly entangled state whose linear entropy is always equal to $1/2$. In particular, if the initial state of the atom is $\vert \pm \rangle$, then the cavity-fields are left in a superposition of the states corresponding to the cases where the initial state of the atom is $\vert e\rangle$ or $\vert g \rangle$.

\begin{widetext}

\begin{eqnarray}
\label{Table10}
&& 
\quad\quad\quad
\mbox{Table 1 (ICO): Final cavity-fields state if initially in $\vert 0,0\rangle$} \nonumber \\
&&
\begin{array}{|c|c|c|c|c|}
\hline
\text{Initial/Final} & \vert 0\rangle_{c} \otimes \vert e \rangle & \vert 0\rangle_{c} \otimes \vert g \rangle & \vert 1\rangle_{c} \otimes \vert e \rangle & \vert 1\rangle_{c} \otimes \vert g \rangle \\ %
\hline
\vert e \rangle & \vert 0,0\rangle & \vert G+\rangle & \mathrm{NP} & \vert G- \rangle \\
\hline
\vert g \rangle & \mathrm{NP} & \vert 0,0\rangle & \mathrm{NP}  & \mathrm{NP}
 \\
\hline
\vert \pm \rangle & \vert 0,0\rangle & d_{07}\vert \phi_{gg}^{0} \rangle \pm d_{08}e^{-i\omega t} \vert G+ \rangle & \mathrm{NP}  &  \vert G- \rangle  \\
\hline
\end{array}
\end{eqnarray}
Table 1 presents the state in which the cavity-fields are prepared given the initial state of the atom in the first column and the results of the measurements of the control qubit and the atom in the first row. NP stands for \textit{not possible} because the probability for such case is zero. The kets $\vert G\pm \rangle$ and $\vert \phi_{gg}^{0} \rangle$ are defined in (\ref{gain}) and Table 2, respectively. The coefficients $d_{07}$ and $d_{08}$ are given in  (\ref{d0j}).

\begin{eqnarray}
\label{Table2}
&& 
\quad \ 
\mbox{Table 2 (ICO): Final cavity-fields state if initially in $\vert n,n\rangle \ \ (n>0)$}  \nonumber \\
&& \ \
\begin{array}{|c|c|c|}
\hline
\text{Initial/Final} & \vert 0 \rangle_{c}\otimes \vert e \rangle & \vert 0 \rangle_{c}\otimes\vert g \rangle \\ %
\hline
\vert e \rangle & \vert \phi_{ee}^{0} \rangle 
=  d_{01}\vert n,n\rangle + d_{02}\vert S+ \rangle  & \vert G+ \rangle \\
\hline
\vert g \rangle & \vert L+ \rangle  & \vert \phi_{gg}^{0} \rangle 
= d_{05}\vert n,n\rangle  + d_{06}\vert S_{+} \rangle 
 \\
\hline
\vert \pm \rangle &  d_{03}\vert \phi_{ee}^{0} \rangle \pm e^{i\omega t}d_{04}\vert L+ \rangle  & d_{07}\vert \phi_{gg}^{0} \rangle  \pm  d_{08}e^{-i\omega t} \vert G+ \rangle\\
\hline
\end{array}
\end{eqnarray}
Table 2 presents the state in which the cavity-fields are prepared given the initial state of the atom in the first column and the results of the measurements of the control qubit ($\vert 0 \rangle_{c}$) and the atom in the first row. The kets are defined in (\ref{swap})-(\ref{lose}) and the coefficients are given in  (\ref{d0j}).

\begin{eqnarray}
\label{Table3}
&& 
\quad \ \
\mbox{Table 3 (ICO): Final cavity-fields state if initially in $\vert n,n\rangle \ \ (n>0)$} \nonumber \\
&& \quad\quad \ \ 
\begin{array}{|c|c|c|}
\hline
\text{Initial/Final} & \vert 1 \rangle_{c} \otimes \vert e \rangle & \vert 1 \rangle_{c} \otimes  \vert g \rangle \\ 
\hline
\vert e \rangle & \vert S-\rangle  & 
\vert G- \rangle 
\\
\hline
\vert g \rangle & \vert L- \rangle  & \vert S- \rangle 
 \\
\hline
\vert \pm \rangle &   d_{11} \vert S- \rangle \pm d_{12}e^{i\omega t}\vert L- \rangle  & d_{13}\vert S- \rangle  \mp  d_{14}e^{-i\omega t}\vert G- \rangle  \\
\hline
\end{array}
\end{eqnarray}
Table 3 presents the state in which the cavity-fields are prepared given the initial state of the atom in the first column and the results of the measurements of the control qubit ($\vert 1 \rangle_{c}$) and the atom in the first row. The kets are defined in (\ref{swap})-(\ref{lose}) and the coefficients are given in  (\ref{d1j}).

\end{widetext}

Now consider the case where the initial state of the cavity-fields is $\vert n,m \rangle$ with $n \not= m$. It is found that the linear entropy associated with the cavity-fields state is independent of $\omega$ and symmetric in $n$ and $m$ (that is, when the linear entropy is viewed as a function of $n$ and $m$, $S_{L}(n,m)$, one has $S_{L}(n,m) = S_{L}(m,n)$ for all the considered initial and final states). In particular, the linear entropy is the same when the initial state of the atom is either $\vert + \rangle$ or $\vert - \rangle$. Furthermore, when $n \not= m$ are both positive integers one can choose transit times $T$ so that the cavity-fields are prepared in a highly entangled state whose linear entropy reaches a maximum value of $1/2$ if the control qubit was found in $\vert 1 \rangle_{c}$. This maximum value is also reached if the control qubit was found in $\vert 0 \rangle_{c}$, the atom was found in $\vert e \rangle$ ($\vert g \rangle$), and the initial state of the atom was $\vert g \rangle$ ($\vert e \rangle$). It can also reach a maximum value of $2/3$ if the control qubit was found in $\vert 0 \rangle_{c}$, the atom was found in the state $\vert e \rangle$ ($\vert g \rangle$), and the initial state of the atom was $\vert e \rangle$ ($\vert g \rangle$). 

%%%%
%%%%
%%%%
\section{Cavities in series}

We now consider the case of cavities in series, see Figs.~\ref{Figure1a} and \ref{Figure1b}, and compare it with the case of ICO. To describe the situation in which the atom first passes through cavity $C_0$ ($C_1$) and then through $C_1$ ($C_0$), one must choose $\theta = 0$ $(\pi/2)$ and $\varphi = 0$ in (\ref{CQDstate}). Hence, the state of the atom-cavity-fields at time $t > T_{1} + T$ when the atom has exited both cavities and in the IP is $\vert C_{i}C_{j}(T)\rangle$ in (\ref{orderc1c0}) or (\ref{orderc0c1}) if the atom first transits through cavity $C_j$ and then through $C_i$. In the rest of this section we only consider the state $\vert C_{1}C_{0}(T) \rangle$ defined in (\ref{orderc1c0}) and (\ref{CS2}). Applying the unitary transformation $U_{I}(t)$ to $\vert C_{1}C_{0}(T) \rangle$, it follows that the state of the atom-cavity-fields in SP is (except for a global phase factor)
\begin{eqnarray}
\label{SeriesPsi}
\vert \aleph (t) \rangle &=& \vert e \rangle\otimes N_{f_{e}}\vert f_{e}(t) \rangle + \vert g \rangle \otimes N_{f_{g}} \vert f_{g}(t) \rangle ,
\end{eqnarray}
where
\begin{eqnarray}
\label{fe}
\vert f_{e}(t) \rangle &=& \frac{1}{N_{f_{e}}}\Bigg[ \left( c_{1}\vert n \rangle_{0}  + c_{2}e^{i\omega t} \vert n-1 \rangle_{0} \right) \otimes \vert m \rangle_{1} \cr
&& + \left( c_{5}e^{i\omega t} \vert n \rangle_{0}  + c_{6} \vert n+1 \rangle_{0} \right) \otimes \vert m - 1 \rangle_{1} \Bigg], \cr
\vert f_{g}(t) \rangle &=& \frac{1}{N_{f_{g}}} \Bigg[ \left( c_{7}e^{i\omega t}\vert n \rangle_{0}  + c_{8} \vert n+1 \rangle_{0} \right) \otimes \vert m \rangle_{1} \cr
&& + \left( c_{3} \vert n \rangle_{0}  + c_{4}e^{i\omega t} \vert n-1 \rangle_{0} \right) \otimes \vert m + 1 \rangle_{1} \Bigg], \quad\quad
\end{eqnarray}
and the normalization constants are
\begin{eqnarray}
\label{fg}
N_{f_{e}}
&=& \sqrt{ \vert c_{1} \vert^{2} + \vert c_{2} \vert^{2} + \vert c_{5} \vert^{2} + \vert c_{6} \vert^{2} } , \cr
N_{f_{g}}
&=& \sqrt{ \vert c_{3} \vert^{2} + \vert c_{4} \vert^{2} + \vert c_{7} \vert^{2} + \vert c_{8} \vert^{2} } .
\end{eqnarray}
In (\ref{fe}) and (\ref{fg}) all the $c_{k}$ are evaluated at $T$. The expression in (\ref{SeriesPsi}) is valid for any initial state of the atom. One must substitute the values of $\chi$ and $\xi$ corresponding to the considered case [see the paragraph after (\ref{Atom})].

We now measure the state of the atom at time $t$ to see if it is found in the excited $\vert e \rangle$ or ground $\vert g \rangle$ state. If the atom was found in the state $\vert e \rangle$ ($\vert g \rangle$), then the state of the cavity-fields immediately after the measurement is $\vert f_{e}(t) \rangle$ ($\vert f_{g}(t) \rangle$). Notice that the linear entropy of the reduced density operators associated with these states is always $\leq 1/2$. This is in contrast with the ICO results of the previous section because it was found that, in some cases, the linear entropy could reach a value of $2/3$.

\begin{widetext}

\begin{eqnarray}
\label{TableS1}
&&
\mbox{Table 4 (Cavities in series): Final cavity-fields state if initially in $\vert 0,0\rangle$} \nonumber \\
&& \quad\quad\quad\quad\quad\quad\quad
\begin{array}{|c|c|c|}
\hline
\text{Initial/Final} & \vert e \rangle & \vert g \rangle \\ 
\hline
\vert e \rangle & \vert 0, 0 \rangle   & 
c_{8}\vert 1, 0 \rangle 
+ c_{3}\vert 0, 1\rangle 
\\
\hline
\vert g \rangle & \mathrm{NP}  & \vert 0, 0\rangle
 \\
\hline
\vert \pm \rangle &   \vert 0,0 \rangle  &  \vert \phi_{g\pm}(t) \rangle  \\
\hline
\end{array}
\end{eqnarray}
Table 4 presents the (possibly nonnormalized) state in which the cavity-fields are prepared for cavities in series given the initial state of the atom in the first column and the result of the measurement of the state of the atom in the first row. The initial state of the cavity-fields is $\vert 0,0\rangle$. The ket $\vert \phi_{g\pm}(t)\rangle$ is defined in (\ref{SeriesPhi}) and the coefficients are given in  (\ref{Cterms}) and must be evaluated at $T$ with the values of $\chi$ and $\xi$ of the considered case [see the paragraph after (\ref{Atom})].
NP stands for \textit{not possible} because the probability for such a case is zero.

\begin{eqnarray}
\label{Table5}
&& 
\mbox{Table 5 (Cavities in series): Final cavity-fields state if initially in $\vert n,n\rangle$ ($n>0$)} \nonumber \\
&&
\quad\quad\ \
\begin{array}{|c|c|c|}
\hline
\text{Initial/Final} & \vert e \rangle & \vert g \rangle \\ 
\hline
\vert e \rangle & c_{1}\vert n, n \rangle 
+ c_{6}\vert n+1, n-1\rangle  & 
c_{8}\vert n+1, n \rangle 
+ c_{3}\vert n, n+1\rangle 
\\
\hline
\vert g \rangle & c_{2}\vert n-1, n \rangle 
+ c_{5}\vert n, n-1\rangle  & c_{7}\vert n, n \rangle 
+ c_{4}\vert n-1, n+1\rangle
 \\
\hline
\vert \pm \rangle &   \vert \phi_{e\pm}(t) \rangle  &  \vert \phi_{g\pm}(t) \rangle  \\
\hline
\end{array}
\end{eqnarray}
Table 5 presents the nonnormalized state in which the cavity-fields are prepared for cavities in series given the initial state of the atom in the first column and the result of the measurement of the state of the atom in the first row. The initial state of the cavity-fields is $\vert n,n\rangle$ with $n>0$. The kets $\vert \phi_{e\pm}(t)\rangle$ and  $\vert \phi_{g\pm}(t)\rangle$ are defined in (\ref{SeriesPhi}) and the coefficients are given in  (\ref{Cterms}) and must be evaluated at $T$ with the values of $\chi$ and $\xi$ of the considered case [see the paragraph after (\ref{Atom})].
\end{widetext}

Tables 4 and 5 present the (possibly) nonnormalized state of the cavity-fields (except for a physically irrelevant global phase factor) in the SP immediately after the measurement of the state of the atom when the initial state of the cavity-fields is $\vert n, n \rangle$ with $n=0$ or $n>0$, respectively. In each table, the first column from left to right presents the initial state of the atom, while the first row from top to bottom shows the state in which the atom was found after the measurement. The nonnormalized states $\vert \phi_{e\pm}(t) \rangle$ and $\vert \phi_{g\pm}(t) \rangle$ are defined by 
\begin{eqnarray}
\label{SeriesPhi}
\vert \phi_{e\pm}(t) \rangle
&=& c_{1}\vert n,n\rangle +  c_{6} \vert n+1,n-1 \rangle \cr
&& + e^{i\omega t} \left[ c_{2}\vert n-1,n\rangle + c_{5} \vert n,n-1 \rangle \right]  , \cr
\vert \phi_{g\pm}(t) \rangle
&=& c_{8}\vert n+1,n\rangle +  c_{3} \vert n,n+1 \rangle \cr
&& + e^{i\omega t} \left[ c_{7}\vert n,n\rangle + c_{4} \vert n-1,n+1 \rangle \right]  .
\end{eqnarray}
All the coefficients must be evaluated at $T$ with the values of $\chi$ and $\xi$ corresponding to the considered case, see the paragraph below (\ref{Atom}). 

\begin{figure}[htbp]
   \centering
   \includegraphics[scale=0.5]{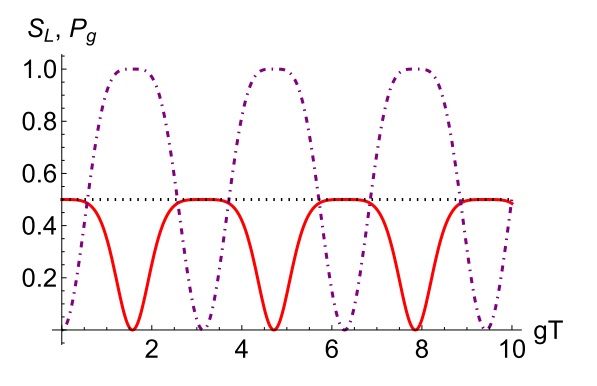}
\caption{The figure illustrates the linear entropy $S_{L}$ (red-solid line) in (\ref{SeriesG}) after the atom has been found in the ground state, as well as the probability $P_{g}$ to find the atom in the ground state (purple-dotdashed line) and the horizontal line $1/2$ (black-dotted line). Initially the atom is in the excited state and the cavity-fields are in their vacuum states.
}
  \label{Figure2}
\end{figure}

It is immediately apparent that the state of the cavity-fields can present entanglement for some values of the transit time $T$. The states in Table 4 should be compared with those of Table 1 for the same initial and final conditions, while the ones in Table 5 should be compared with the corresponding states in Tables 2 and 3. The main difference is that ICO can always lead to maximally entangled states, while for cavities in series one has to adjust the transit time $T$ to special values in order to obtain highly entangled cavity-fields states but the probability of obtaining those states can be very small. An explicit example that illustrates this well is obtained by considering that the atom is found in the ground state, that the atom is initially in the excited state, and that the cavity-fields are initially in their vacuum states. In the case of ICO, from Table 1 the cavity-fields are prepared in the highly entangled one-photon NOON states $\vert G \pm \rangle$. In the case of cavities in series, it is straightforward to show that the linear entropy is 
\begin{eqnarray}
\label{SeriesG}
S_{L}(gT) = 1 - \mathrm{Tr}_{0}[\rho_{0}^{2}(t)] = \frac{8 \mathrm{cos}^{2}(gT)}{[3+\mathrm{cos}(2gT)]^{2}},
\end{eqnarray}
where $\rho_{0}(t) = \mathrm{Tr}_{1}\left[ \vert f_{g}(t) \rangle\langle f_{g}(t) \vert \right]$ and $\mathrm{Tr}_{j}$ is the trace over the degrees of freedom of the cavity-field in $C_{j}$. $S_{L}(gT)$  reaches a maximum value $1/2$ when $gT = N\pi$ with $N$ a positive integer. Nevertheless, the probability $P_{g}$ of finding the atom in the ground state given the initial conditions that we are considering is zero for these values of $T$, so it is difficult to obtain a cavity-fields state whose associated linear entropy is close to $1/2$. This is illustrated in Fig.~\ref{Figure2} which shows the linear entropy (red-solid line) in (\ref{SeriesG}) and the probability $P_{g}$ (purple-dotdashed line) of finding the atom in the ground state. 

Inspecting Tables 1-5 observe that with cavities in series one cannot prepare the cavity-fields in the $\vert S\pm \rangle$ states where the \textit{cavity-fields interchange one photon}. However, this can be accomplished with ICO and, in addition, there are also cases where the atom can be found in same state as the initial one (see Tables 2 and 3). 

\begin{figure}[htbp]
   \centering
   \includegraphics[scale=0.5]{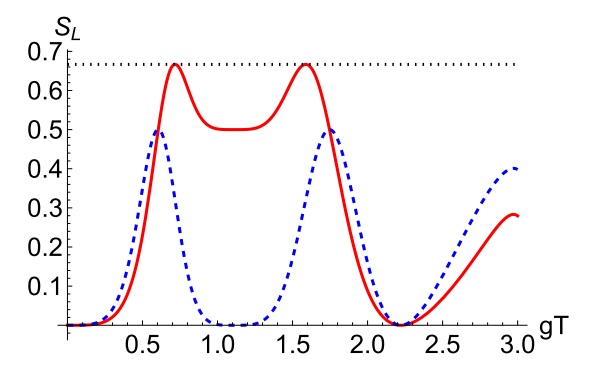}
\caption{The figure illustrates the linear entropy $S_{L}$ associated with the state of cavity-fields in the cavities in series case (blue-dashed line) and when ICO is present (red-solid line). It considers that the atom is found in the excited state and that initially the atom was in the excited state and the cavity-fields were in the state $\vert 1,1 \rangle$. When ICO is present it is assumed that the control qubit was found in the state $\vert 0 \rangle_{c}$. The horizontal line $2/3$ also appears (black-dotted line).
}
  \label{Figure4}
\end{figure}

One can restrict oneself to the cases where the cavity-fields are not always prepared in a maximally entangled state when ICO is present. For some of these, one can choose transit times $T$ so that the state of the cavity-fields is more entangled in the cavity in series case than when ICO is present. This is illustrated in Fig.~\ref{Figure4}, which considers that the atom is found in the excited state and that initially the atom was in the excited state and the cavity-fields were in the state $\vert 1,1\rangle$. For ICO it is assumed that the control qubit was found in the state $\vert 0 \rangle_{c}$ because otherwise the cavity-fields are always left in the maximally entangled state $\vert S-\rangle$. The figure shows the linear entropies of the caviy in series (blue-dashed line) and ICO (red-solid line) cases. Notice that there are values of $gT$ such that the graph of the cavity in series case is above the graph of the ICO case. 

Finally, we make some brief comments about the case where the cavity-fields are initially in the state $\vert n,m\rangle$ with $n\not= m$. There are situations where one can adjust the transit time so that the linear entropy of the cavity-fields state after the measurement of the state of the atom is $1/2$. For example, this happens when the initial state of the atom-cavity-fields is $\vert e \rangle\otimes \vert 0,m \rangle$ with $m>0$ and the atom is found in either the excited or ground state.

%%%%
%%%%
%%%%
 
\subsection{Rabi oscillations}

\begin{figure}[htbp]
   \centering
   \includegraphics[scale=0.5]{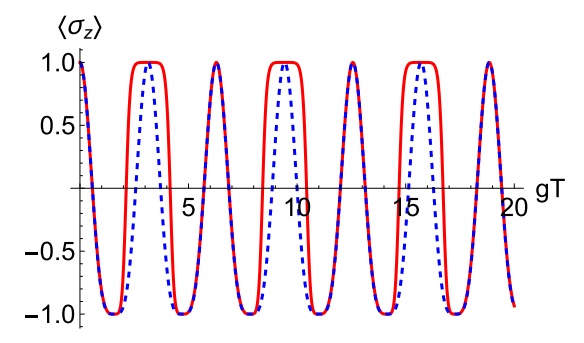}
\caption{Comparison of $\langle \sigma_{z} \rangle$ as a function of $gT$ for cavities in series (blue-dashed line) and for ICO  (red-solid line) when the atom and the cavity-fields are initially in the excited and vacuum states, respectively. In the case of ICO the control qubit was found in $\vert 0 \rangle_{c}$.}
  \label{Figure3}
\end{figure}

In the Jaynes-Cummings model, if the atom is initially in its excited state and the cavity field is in a Fock state, then $\langle \sigma_{z} \rangle$ exhibits the well-known sinusoidal Rabi oscillations \cite{gerry2023introductory}.  The purpose of this section is to determine what happens to $\langle \sigma_{z} \rangle$ when one has two cavities in series and when there is ICO. To determine the behavior of the Rabi oscillations we vary the transit time $T$, that is, the time it takes the atom to transit each cavity. We only consider the case where initially the atom is in the excited state and each cavity-field is in the vacuum state.

 For two cavities in series where the atom first passes through $C_{0}$ and then through $C_{1}$, one gets 
\begin{eqnarray}
\label{RabiSeries}
\langle C_1 C_0 (T)  \vert \sigma_{z} \vert C_1 C_0 (T)\rangle 
&=& \text{cos}(2 gT) \text{cos}^{2}(gT) - \mathrm{sin}^{2}(gT) . \cr
&&
\end{eqnarray}
In the case of ICO, using (\ref{CQDfinal}) one finds that
\begin{eqnarray}
\label{RabiICO}
\langle \psi_{0}(t) \vert \sigma_{z} \vert \psi_{0}(t)\rangle 
&=& \frac{1}{4 \mathcal{N}_{0}^2}\Big\{ \vert c_1 + s_1 \vert^2 + \vert c_6 \vert^2  + \vert s_6 \vert^2 \cr
&& -\vert c_3+s_8 \vert^2  -\vert c_8+s_3 \vert^2 \Big\} .
\end{eqnarray}
In (\ref{RabiICO}) the $c_{k}$ and $s_{k}$ are evaluated at time $T$ with $\xi = \chi = 0$. Also, $
\langle \psi_{1}(t) \vert \sigma_{z} \vert \psi_{1}(t)\rangle = -1$ whenever $gT \not= N\pi$ with $N$ a positive integer. This last condition is necessary to have $P_{1\vert e} >0$, see (\ref{ProbControl}). 

Fig.~\ref{Figure3} compares (\ref{RabiSeries}) (blue-dashed line) with  (\ref{RabiICO}) (red-solid line). In the case of ICO the Rabi oscillations no longer have a sinusoidal form. They disappear completely if the control qubit is found in $\vert 1 \rangle_{c}$ and \textit{plateaus} appear when the control qubit is found in the state $\vert 0 \rangle_{c}$. 

%%%%
%%%%
%%%%
\section{Conclusions}

We have introduced indefinite causal order (ICO) into the paradigm of atom-field interactions and investigated its influence on the atom-field observables. As a first approach, we modeled the interaction between one atom and two cavity-fields, each one initially in a Fock state, using the Jaynes-Cummings model. The results can be extended to more scenarios, for example, considering coherent states or squeezed states or extending the approach to more complex models of interaction, such as the Tavis-Cummings model where $N$ atoms are coupled to a single-mode cavity-field. 

Our results suggest that ICO can offer a new way to coherently control the interaction between the electromagnetic field of two cavities and one atom. This new control allows us, for example, to create entanglement between two electromagnetic fields that never interact directly. Depending on the result of the measurement of the state of the control qubit, the cavity-fields can be prepared in a maximally entangled \textit{Bell-like} state independent of both the initial (excited or ground) state of the atom and the time $T$ it takes the atom to transit each cavity. Adjusting $T$ one can guarantee that the cavity-fields are always prepared in a maximally entangled state that can take the form of one- or two-photon NOON states. The main difference with the case of cavities in series is that ICO can always lead to maximally entangled states, while for cavities in series one has to adjust the transit time $T$ to special values in order to obtain highly entangled cavity-fields states and the probability of obtaining those states can be very small. We also show that ICO can interchange one photon between both cavities without changing the state of the atom, something that is impossible to achieve for two cavities in well-defined order. Furthermore, the vacuum Rabi oscillations either disappear or lose their sinusoidal form with ICO.

ICO in cavity quantum electrodynamics (cQED) can in principle be implemented using a trapped-atom Sagnac interferometer where the atomic wave packet is split and recombined along circular trajectories \cite{Sagna2020, Sagna2021, Sagna2024}. In this case, the cavities can be placed along the circular trajectories. In this way, the atom traveling along those trajectories will interact with the two cavity fields in a superposition of different orders. Our work can be seen as an effort to explore new research avenues to investigate ICO on different platforms and to look for new physical effects that can be used for the manipulation of atom-field observables.

%%%%
%%%%
%%%%
\section*{Appendix}

The coefficients appearing in the expressions for $\vert C_{1}C_{0}(\tau)\rangle$ and $\vert C_{0}C_{1}(\tau)\rangle$ are given by
\begin{eqnarray}
	\label{Cterms}
	c_1 (\tau) &=&\mathrm{cos}(\xi)\mathrm{cos}(\gamma_n T) \mathrm{cos}(\gamma_m \tau), \cr
	c_2 (\tau)&=&-i e^{i \chi} \mathrm{sin}(\xi)\mathrm{sin}(\gamma_{n-1} T) \mathrm{cos}(\gamma_m \tau), \cr
	c_3 (\tau)&=&-i \mathrm{cos}(\xi)\mathrm{cos}(\gamma_n T) \mathrm{sin}(\gamma_m \tau), \cr
	c_4 (\tau)&=&-e^{i\chi} \mathrm{sin}(\xi) \mathrm{sin}(\gamma_{n-1} T) \mathrm{sin}(\gamma_m \tau), \cr
	c_5 (\tau)&=& -i e^{i\chi} \mathrm{sin}(\xi) \mathrm{cos}(\gamma_{n-1} T) \mathrm{sin}(\gamma_{m-1} \tau), \cr
	c_6 (\tau)&=& -\mathrm{cos}(\xi) \mathrm{sin}(\gamma_n T) \mathrm{sin}(\gamma_{m-1} \tau), \cr
	c_7 (\tau)&=& e^{i\chi} \mathrm{sin}(\xi) \mathrm{cos}(\gamma_{n-1} T) \mathrm{cos}(\gamma_{m-1} \tau), \cr
	c_8 (\tau)&=&-i\mathrm{cos}(\xi) \mathrm{sin}(\gamma_n T) \mathrm{cos}(\gamma_{m-1} \tau), 
\end{eqnarray}
and
\begin{eqnarray}
	\label{Sterms}
	s_1(\tau) &=&\mathrm{cos}(\xi)\mathrm{cos}(\gamma_m T) \mathrm{cos}(\gamma_n \tau), \cr
	s_2(\tau) &=&-i e^{i \chi} \mathrm{sin}(\xi) \mathrm{sin}(\gamma_{m-1} T) \mathrm{cos}(\gamma_n \tau), \cr
	s_3(\tau) &=&-i \mathrm{cos}(\xi)\mathrm{cos}(\gamma_m T) \mathrm{sin}(\gamma_n \tau), \cr
	s_4(\tau) &=&-e^{i\chi} \mathrm{sin}(\xi) \mathrm{sin}(\gamma_{m-1} T) \mathrm{sin}(\gamma_n \tau), \cr
	s_5(\tau) &=& -i e^{i\chi} \mathrm{sin}(\xi) \mathrm{cos}(\gamma_{m-1} T) \mathrm{sin}(\gamma_{n-1} \tau), \cr
	s_6(\tau) &=& -\mathrm{cos}(\xi) \mathrm{sin}(\gamma_m T) \mathrm{sin}(\gamma_{n-1} \tau), \cr
	s_7 (\tau)&=& e^{i\chi} \mathrm{sin}(\xi) \mathrm{cos}(\gamma_{m-1} T) \mathrm{cos}(\gamma_{n-1} \tau), \cr
	s_8(\tau) &=&-i\mathrm{cos}(\xi) \mathrm{sin}(\gamma_m T) \mathrm{cos}(\gamma_{n-1} \tau),
\end{eqnarray}

\noindent where $\gamma_{n}=g \sqrt{n+1}$.

The normalization factors appearing in (\ref{EdoCampos}) and (\ref{EdoCamposb}) are given by $(j=0,1)$
\begin{eqnarray}
\label{Nlj}  
\mathcal{N}_{ej}^2 
&=& \vert c_{1} + (-1)^{j}s_{1} \vert^{2} + \vert c_{2} + (-1)^{j}s_{5} \vert^{2} \cr
&& + \vert c_{5} + (-1)^{j}s_{2} \vert^{2} + \vert c_{6} \vert^{2} + \vert s_{6} \vert^{2} , \cr
\mathcal{N}_{gj}^2 
&=& \vert c_{7} + (-1)^{j}s_{7} \vert^{2} + \vert c_{8} + (-1)^{j}s_{3} \vert^{2} \cr
&& + \vert c_{3} + (-1)^{j}s_{8} \vert^{2} + \vert c_{4} \vert^{2} + \vert s_{4} \vert^{2} , 
\end{eqnarray}
where the $c_{k}$ and $s_{k}$ are all evaluated at $T$.

We now give expressions for the coefficients $d_{ij}$ appearing in Tables 2 and 3.

Define
\begin{eqnarray}
\label{abdc}
a_{n} &=& - \mathrm{sin}(gT\sqrt{n}) \mathrm{sin}(gT\sqrt{n+1}) , \cr
b_{n,m} &=& i \mathrm{sin}(gT\sqrt{m}) \left[ \mathrm{cos}(gT\sqrt{n}) - \mathrm{cos}(gT\sqrt{n+1}) \right] , \cr
B_{n,m} &=& -i\mathrm{sin}(gT\sqrt{m}) \left[ \mathrm{cos}(gT\sqrt{n}) + \mathrm{cos}(gT\sqrt{n+1}) \right] , \cr
c_{n} &=& 2 \mathrm{cos}^{2}(gT\sqrt{n}) .
\end{eqnarray}
Then
\begin{eqnarray}
\label{d0j}
d_{01} &=& \frac{c_{n+1}}{\sqrt{ \vert c_{n+1}  \vert^{2} + 2\vert a_{n} \vert^{2} } } , \quad d_{02} = \frac{\sqrt{2} a_{n}}{\sqrt{ \vert c_{n+1}  \vert^{2} + 2\vert a_{n} \vert^{2} } } , \cr
d_{03} &=& \frac{\sqrt{\vert c_{n+1} \vert^{2} + 2 \vert a_{n} \vert^{2}}}{\sqrt{ \vert c_{n+1}  \vert^{2} + 2 \vert a_{n} \vert^{2} + 2\vert B_{n,n} \vert^{2} } } , \cr
d_{04} &=& \frac{\sqrt{2}B_{n,n}}{\sqrt{ \vert c_{n+1}  \vert^{2} + 2 \vert a_{n} \vert^{2} + 2\vert B_{n,n} \vert^{2} } } , \cr
d_{05} &=& \frac{c_{n}}{\sqrt{ \vert c_{n}  \vert^{2} + 2 \vert a_{n} \vert^{2}} } , \quad d_{06} = \frac{\sqrt{2} a_{n}}{\sqrt{ \vert c_{n}  \vert^{2} + 2 \vert a_{n} \vert^{2}} } ,\cr
d_{07} &=& \frac{\sqrt{\vert c_{n} \vert^{2} + 2 \vert a_{n} \vert^{2}}}{\sqrt{ \vert c_{n}  \vert^{2} + 2 \vert a_{n} \vert^{2} + 2\vert B_{n,n+1} \vert^{2} } } , \cr
d_{08} &=& \frac{\sqrt{2} B_{n,n+1}}{\sqrt{ \vert c_{n}  \vert^{2} + 2 \vert a_{n} \vert^{2} + 2\vert B_{n,n+1} \vert^{2} } } ,
\end{eqnarray}
and
\begin{eqnarray}
\label{d1j}
d_{11} &=& \frac{a_{n}}{\sqrt{\vert a_{n} \vert^{2} + \vert b_{n,n} \vert^{2}}} , \quad\quad d_{12} = \frac{b_{n,n}}{\sqrt{\vert a_{n} \vert^{2} + \vert b_{n,n} \vert^{2}}} , \cr
d_{13} &=& \frac{a_{n}}{\sqrt{\vert a_{n} \vert^{2} + \vert b_{n,n+1} \vert^{2} }} , \quad d_{14} = \frac{b_{n,n+1}}{\sqrt{\vert a_{n} \vert^{2} + \vert b_{n,n+1} \vert^{2} }} . \cr
&&
\end{eqnarray}

Finally, we present the probabilities for the various possible results. Let $P_{t}(l'\vert j,l)$ denote the probability to find the atom in the state $\vert l'\rangle$ at time $t > T_{1} + T$ given that the control qubit was found in the state $\vert j \rangle_{c}$ and that the initial state of the atom was $\vert l \rangle$. Here $l',l = e$, $g$, and $\pm$ and $j = 0,1$. Also, the initial state of the cavity-fields is $\vert n, m \rangle$. Then, $P_{t}(l'\vert j,l) = p(l'\vert j,l)/(4\mathcal{N}_{j}^{2})$ with
\begin{eqnarray}
\label{PC}
p(e \vert 1,e) &=& \mathrm{sin}^{2}(gT\sqrt{n})\mathrm{sin}^{2}(gT\sqrt{m+1}) \cr
&& + \mathrm{sin}^{2}(gT\sqrt{n+1})\mathrm{sin}^{2}(gT\sqrt{m}) , \cr
p(e \vert 1,g) &=& \vert b_{n,m} \vert^{2} + \vert b_{m,n} \vert^{2} , \cr
p(e \vert 1,\pm) &=& \frac{1}{2}\left[ p(e \vert 1,e) + p(e \vert 1,g)\right] , \cr
p(g \vert 1,e) &=& \vert b_{n,m+1} \vert^{2} + \vert b_{m,n+1} \vert^{2} , \cr
p(g \vert 1,g) &=& p(e\vert 1,e), \cr
p(g \vert 1,\pm) &=& \frac{1}{2} \left[ p(g\vert 1,e) + p(g\vert 1,g)  \right] , \cr
p(e \vert 0,e) &=& c_{n+1}c_{m+1} + p(e \vert 1,e) , \cr
p(e \vert 0,g) &=& \vert B_{n,m} \vert^{2} + \vert B_{m,n} \vert^{2} , \cr
p(e\vert 0 , \pm)
&=& \frac{1}{2}\left[ p(e\vert 0 , g) + p(e\vert 0 , e)  \right] , \cr
p(g\vert 0 , e)
&=& \vert B_{n,m+1} \vert^{2} + \vert B_{m,n+1} \vert^{2} , \cr
p(g\vert 0 , g)
&=& c_{n}c_{m} + p(e\vert 1,e), \cr
p(g\vert 0 , \pm)
&=& \frac{1}{2}\left[ p(g\vert 0 , e) + p(g\vert 0 , g)  \right].
\end{eqnarray}
In particular, if $n=m=0$, then 
\begin{itemize}
\item $p(e \vert 1,e) = p(e \vert 1,g) = p(e \vert 1,\pm) = p(e \vert 0,g) = p(g \vert 1,g) = 0$.
\item $p(g \vert 1,e) = 2\mathrm{sin}^{2}(gT)\left[ 1 - \mathrm{cos}(gT) \right]^{2}$.
\item $p(e \vert 0,e) = 4\mathrm{cos}^{4}(gT)$.
\item $p(e \vert 0,\pm) = 2\mathrm{cos}^{4}(gT)$.
\item $p(g \vert 0,e) = 2\mathrm{sin}^{2}(gT)[1 + \mathrm{cos}(gT)]^{2}$.
\item $p(g \vert 0,g) = 4$.
\end{itemize}

%\bibliography{JC_refs}

%

\end{document}